\documentstyle[12pt]{article}
\newcommand{\be}{\begin{equation}}
\newcommand{\ee}{\end{equation}}
\newcommand{\ba}{\begin{eqnarray}}
\newcommand{\ea}{\end{eqnarray}}

\begin{document}
\begin{center}
{\bf HIGHER  ORDER  MATRIX  SUSY  TRANSFORMATIONS  IN
TWO-DIMENSIONAL  QUANTUM  MECHANICS}\\
\vspace{0.3cm}

{\large \bf F.Cannata$^{1,}$\footnote{E-mail: cannata@bo.infn.it}, M.V.Ioffe$^{2,}$\footnote{{\it
Corresponding author.}
Phone:+7(812)4284553; FAX: +7(812)4287240; E-mail:
m.ioffe@pobox.spbu.ru}, A.I.Neelov$^{2,3,}$\footnote{E-mail:
alexey.neelov@unibas.ch}, D.N.Nishnianidze}$^{2,4,}$\footnote{E-mail: qutaisi@hotmail.com}\\
\vspace{0.2cm}
$^1$ Dipartimento di Fisica and INFN, Via Irnerio 46, 40126 Bologna, Italy.\\
$^2$ Sankt-Petersburg State University, 198504 Sankt-Petersburg, Russia\\
$^3$ Institute of Physics, University of Basel, CH-4056 Basel,
Switzerland\\
$^4$ Kutaisi Technical University, 4614 Kutaisi, Republic of
Georgia
\end{center}
\vspace{0.2cm}
\hspace*{0.5in}
\begin{minipage}{5.0in}
{\small
The iteration procedure of supersymmetric transformations
for the two-dimensional Schr\"odinger operator is implemented
by means of the matrix form of factorization
in terms of matrix $2\times 2$ supercharges.
Two different types of iterations are investigated in detail.
The particular case of diagonal initial Hamiltonian is considered,
and the existence of solutions is demonstrated. Explicit examples
illustrate the construction.\\
}
\end{minipage}
\vspace*{0.2cm}
\section*{\bf 1. \quad Introduction}
\vspace*{0.2cm}
\hspace*{3ex}
Supersymmetric Quantum Mechanics (SUSY QM) \cite{witten}, \cite{review}
has become a well-known powerful tool for the investigation of
problems of modern non-relativistic Quantum Mechanics. The main
achievements of SUSY QM during the last two decades involved
isospectral pairs of one-dimensional systems.
In particular, the conventional SUSY QM (with supercharges of first order
in derivatives)
was generalized \cite{ais}, \cite{acdi} to Higher order SUSY QM (HSUSY QM)
with supercharges
$ n-$th order polynomials in momentum. This generalization induces
the deformation of SUSY QM algebra: the anticommutator of
supercharges is now a polynomial in the Superhamiltonian.
The class of pairs of isospectral systems which are
connected (intertwined) by components of supercharge is thus enlarged \cite{acdi}.
Two classes of second order supercharges were
found: {\bf reducible}, which can be written as a chain of
consecutive first order transformations, and {\bf irreducible} second
order transformations. In both cases the intertwined Hamiltonians are
expressed in terms of only one arbitrary function \cite{acdi}.
For a (admittedly nonexhaustive) list of references devoted to
one-dimensional HSUSY QM see, in particular, \cite{HSUSY}.

It is also useful to recall the equivalent terminology
related to the classic mathematical paper by G.Darboux \cite{darboux}.
The conventional
first order SUSY transformations coincide with the so-called Darboux
transformations, which are very useful in Mathematical Physics (see,
for example, \cite{calogero} for nonlinear evolution equations).
Then $n-$th order HSUSY transformations correspond
to the well-known Crum-Krein transformations \cite{crum},
where the eigenfunctions of $ n$-th iterated Hamiltonian are obtained in
terms of $ n $ given eigenfunctions of the initial Hamiltonian by means
of "determinant expressions".
The second iterations of SUSY transformations for matrix one-dimensional
potentials were
considered in \cite{matrix}, recently the general form of matrix Crum-Krein
transformations was constructed \cite{samsonov}.

Much less attention was paid in the literature to {\bf SUSY QM of
two-dimensional systems}. Here we list some partial achievements which have been
obtained in this field.
The direct multi-dimensional generalization of the standard one-dimensional
SUSY QM was elaborated
in \cite{abi}, \cite{abei} for arbitrary space dimensions.
In the simplest two-dimensional case
the supersymmetrization of a given (scalar)
Schr\"odinger Hamiltonian involves not only a second
scalar Hamiltonian but also a matrix $ 2\times 2 $ Schr\"odinger operator.
The spectra
of these three components of the Superhamiltonian are
interrelated, and their wave functions are connected by the components of
supercharges. This approach was used successfully \cite{pauli1},
\cite{pauli2} to find the spectrum and the wave functions of Pauli operators
describing spin $ 1/2 $ particles (with arbitrary values of the gyromagnetic
ratio) in a broad class of non-homogeneous external electromagnetic fields
in terms of spectra and eigenfunctions of two scalar Hamiltonians.

In the series of papers \cite{david} direct intertwining
relations between a pair of scalar two-dimensional Hamiltonians were
studied. A wide variety of particular solutions was found, leading to new
integrable quantum systems which are not amenable to separation of variables.
This approach allowed to introduce \cite{two} two new methods for the study of
the two-dimensional Schr\"odinger equation: $ SUSY-$separation of
variables and shape invariance. Multi-dimensional SUSY QM for an arbitrary oriented
Riemannian manifolds was studied in \cite{kamran}.
In a very recent paper \cite{mateos}
the Hamilton-Jacobi approach of Classical Mechanics was applied to study
classical two-dimensional supersymmetric models and their
quantization.

The present paper addresses the problem of the construction
of a chain of consecutive supersymmetric
(generalized Darboux) transformations for two-dimensional Schr\"odinger
systems allowing for a deeper insight into isospectrality in multidimensional cases.
Let us remark that the only early attempt to
construct such iterations failed (see Section 4 of the first paper in \cite{abi}).
Even the very possibility of making iterations for two-dimensional systems remained a
task to be solved.

The new matrix supersymmetric factorization of the two-dimensional Pauli operator,
recently proposed in \cite{pauli2}, can provide a new chance
to construct iterations of first order (reducible) supersymmetric transformations.
Indeed, in contrast to the method of quasifactorization \cite{abi}, where the vector components of
supercharges include scalar superpotential, the matrix factorization of \cite{pauli2} involves
the components of supercharges with matrix superpotentials. Therefore all elements of the SUSY intertwining
relations are now represented by $2\times 2$ matrix differential operators.

The organization of the paper is the following. In Section 2 the chain of
of SUSY transformations is performed by
gluing the corresponding components of two successive
Superhamiltonians. The generalization of this prescription includes
an additional global unitary transformation (Section 3).
In Section 4 we analyze the particularly interesting
case when the initial Hamiltonian is diagonal, and one
supersymmetrizes a scalar Hamiltonian by building partner
Hamiltonians via second order SUSY transformations. This construction,
in reverse order, can
be also used for diagonalization of a class of two-dimensional
$ 2\times 2 $ matrix Schr\"odinger operators or for matrix systems
of coupled second order differential equations.

\section*{\bf 2.\quad Second iteration by gluing}
\vspace*{0.2cm}
\hspace*{3ex} Before introducing the new form of SUSY factorization
for two-dimensional systems, proposed in \cite{pauli2}, let us start
from the main formulas for the conventional \cite{review}
one-dimensional SUSY QM with partner Hamiltonians and their
eigenfunctions:
\be
H^{(0)} = q^+q^- = -\partial^2 + V^{(0)}(x);\quad
H^{(0)} \Psi_n^{(0)}(x) = E_n \Psi_n^{(0)}(x);\label{h1}
\ee
\be
H^{(1)} = q^-q^+ = -\partial^2 + V^{(1)}(x);\quad
H^{(1)} \Psi_n^{(1)}(x) = E_n \Psi_n^{(1)}(x).\label{h2}
\ee
The operators $ q^{\pm}$ are defined as:
\ba
q^+ = -\partial + W(x);\quad
q^- = (q^+)^{\dagger} = + \partial + W(x);\quad
\partial \equiv d/dx ,
\nonumber
\ea
and they intertwine the components (\ref{h1}), (\ref{h2}) of the
Superhamiltonian:
\be
H^{(0)}q^+ = q^+H^{(1)};\quad
q^- H^{(0)} = H^{(1)}q^- .\label{intertw}
\ee
Then, up to normalization factors,
\be
\Psi_n^{(1)}(x) = q^- \Psi_n^{(0)}(x);\quad
\Psi_n^{(0)}(x) = q^+ \Psi_n^{(1)}(x).\label{psi}
\ee
The intertwinings (\ref{intertw}) are the most important relations of the SUSY QM algebra:
\be
\{\hat Q^+,\hat Q^-\} = \hat H;\quad (\hat Q^+)^2 =(\hat Q^-)^2 = 0;
\quad [\hat H ,\hat Q^{\pm}] =0,
\label{algebra}
\ee
where the Superhamiltonian $ \hat H $ and supercharges
$ \hat Q^{\pm} $ are:
\ba
\hat H =
\left( \begin{array}{cc}H^{(0)}&0\\
0& H^{(1)}
\end{array} \right); \qquad
\hat Q^+ =(\hat Q^-)^{\dagger} = \left( \begin{array}{cc}
0&0\\
q^-&0
\end{array} \right).
\label{definition}
\ea
This procedure of supersymmetric (Darboux) transformations from
$ H^{(0)}, \Psi_{n}^{(0)}(x), E_{n} $ to
$ H^{(1)}, \Psi_{n}^{(1)}(x), E_{n} $ was iterated \cite{acdi}
by reducible supercharges of second order in
derivatives. This iteration leads to a polynomial deformation of
SUSY algebra.

A two-dimensional generalization of standard SUSY QM \cite{abi}
organized (in the Superhamiltonian)
one matrix $2\times 2$ and two scalar Hamiltonians\footnote{For
the case of an arbitrary dimensionality $d$ of coordinate space the Superhamiltonian
contains $(d+1)$ components $H^{(n)}, n=0,1,...,d$ with {\bf matrix}
dimensionality $C_d^n,$ i.e. dimensionality of the space with definite
fermionic occupation number $n$ (see details in \cite{abei}).}.
Each scalar Hamiltonian is separately intertwined with the matrix Hamiltonian,
but the spectra of the two scalar Hamiltonians are not related.
This construction was based on the quasi-factorization of all three
components of the Superhamiltonian in terms of the first order
differential operators $ q_{l}^{\pm}=\mp\partial_l + (\partial_l\chi(x)) $
and\footnote{The summation over repeated indices $ (i=1,2) $ is implied
here and below.} $ p_{l}^{\pm}=\epsilon_{lk}q_{k}^{\mp};\, l,k=1,2.$
A similar quasi-factorization with new supercharges, used
in \cite{abi} to build an iteration of SUSY transformations,
turned out to
be unsuccessful since several necessary constraints were
too complicated to be implemented \cite{abi}. In addition,
it led to a growth of the matrix dimensionality of the new Hamiltonians.

In the present paper we use the method \cite{pauli2} based on
the matrix factorization of components of the two-dimensional Hermitean
Superhamiltonian\footnote{Slightly different notations are chosen here
in comparison with \cite{pauli2}.}:
\ba
H^{(0)}&=&q^{+}q^{-}=
-\Delta^{(2)}+(W_{i})^2 + \sigma_{1}(\partial_{2}W_{1}-\partial_{1}W_{2})-
\sigma_3(\partial_{1}W_{1}+\partial_{2}W_{2}) \label{fac1}\\
H^{(1)}&=&q^{-}q^{+}=
-\Delta^{(2)}+(W_{i})^2 + \sigma_{1}(\partial_{2}W_{1}+\partial_{1}W_{2})+
\sigma_3(\partial_{1}W_{1}-\partial_{2}W_{2}) \label{fac2}\\
& &\partial_{i}\equiv\frac{\partial}{\partial x_{i}} \quad (i=1,2)\quad
\Delta\equiv\partial_{1}^{2}+\partial_{2}^{2}\nonumber
\ea
The $ 2\times 2 $ components  $ q^{\pm}$  of $ 4\times 4 $ supercharge
$ \hat Q^{\pm} $ (see Eq.(\ref{definition})) are
matrix differential operators of first order in derivatives
\footnote{An analogous form of matrix supercharges was used also
in \cite{para} in the context of ParaSUSY QM.}:
\be
q^\pm=\mp\partial_1-i\sigma_2\partial_2+\sigma_1 W_{2} +\sigma_3 W_{1}
\qquad q^+=(q^-)^\dagger , \label{q}
\ee
where $W_{1,2}$ are real, and $ \sigma_{i}\,\,(i=1,2,3) $ are standard
Pauli matrices.
By construction, these operators intertwine the Hamiltonians
(\ref{fac1}), (\ref{fac2})
according to the standard relations (\ref{intertw}) of SUSY algebra.
The relations (\ref{psi}) - (\ref{definition}) of one-dimensional SUSY QM
are satisfied as well, with the proviso that we deal now with
$ 2\times 2 $ {\bf matrix} differential operators
$ q^{\pm}, H^{(0)}, H^{(1)} $
and two-component wave functions $ \Psi_{n}^{(0)}, \Psi_{n}^{(1)}.$
Isospectrality of the (matrix) Hamiltonians $H^{(0)},H^{(1)}$ holds,
as usual, except for the possible zero modes of $ q^{\pm} $.

To iterate supersymmetric transformations we impose
the ladder equation (see \cite{ais}, \cite{acdi}) thus gluing
the lower component $ H^{(1)} $ of the Superhamiltonian $ \hat H $
with the upper component $ \widetilde H^{(0)} $ of the next Superhamiltonian
$ \hat{\widetilde H},$ defined by (\ref{fac1}), (\ref{fac2}) with $W$ replaced
by $\widetilde W:$
\be
H^{(1)}=\widetilde H^{(0)} + C,\qquad C = const (real).
\label{gluing}
\ee
Then the second order supercharge $Q^{+}\equiv q^+\tilde q^+$ gives the
following intertwining relations:
\ba
H^{(0)} Q^+ = Q^+ (\widetilde H^{(1)}+C).
\nonumber
\ea
The conditions (\ref{gluing}) can be rewritten as a nonlinear system of differential equations:
\ba
W_{i}^{2}&=&\widetilde W_{i}^{2}+C; \label{nonlinear}\\
\partial_{2}W_{1}&+&\partial_{1}W_{2}=\partial_{2}\widetilde W_{1} -
\partial_{1}\widetilde W_{2}; \label{first}\\
\partial_{1}W_{1}-\partial_{2}W_{2}&=&-\partial_{1}\widetilde W_{1} -
\partial_{2}\widetilde W_{2}. \label{second}
\ea
It may be convenient to express these ladder equations in terms of
complex function of two mutually conjugated complex variables
\be
W(z, z^{\star}) \equiv W_{1}(z, z^{\star}) + iW_{2}(z, z^{\star});\quad
z \equiv x_{1}+ix_{2}\quad z^{\star} \equiv x_{1}-ix_{2}, \label{define}
\ee
as follows:
\ba
|W(z, z^{\star})|^{2}&=&|\widetilde W(z, z^{\star})|^{2} + C; \label{1}\\
\bar\partial W(z, z^{\star})&=&-\partial \widetilde W(z, z^{\star});
\label{2}\\
\partial \equiv \frac{\partial}{\partial z}&=&\frac{1}{2}(\partial_{1}
-i\partial_{2});
\quad
\bar\partial\equiv \frac{\partial}{\partial z^{\star}}=
\frac{1}{2}(\partial_{1}+i\partial_{2}). \nonumber
\ea
The general solution of Eq.(\ref{2}) can be written by means of an
{\bf arbitrary} complex function:
\be
W(z, z^{\star})=-\partial F(z, z^{\star});\quad \widetilde W(z,
z^{\star})=\bar\partial F(z, z^{\star}); \quad F(z, z^{\star})
\equiv F_1(z, z^{\star})+iF_2(z, z^{\star}) \label{F}
\ee
with real $ F_1 $ and $ F_2.$

In terms of $F_1(z,z^{\star}), F_2(z,z^{\star})$  Eq.(\ref{1}) reads:
\be
(\partial_1F_1)(\partial_2F_2) - (\partial_2F_1)(\partial_1F_2) = C.
\label{FF}
\ee
Its general solution is a general solution of the homogeneous equation summed to a particular
solution of the inhomogeneous one. One can check that the general solution of
the homogeneous equation requires that
$ F_2(x_{1},x_{2}) $ depends on its arguments
via $ F_1(x_{1},x_{2}) $ only:
\be
F_2(x_{1},x_{2})= \Phi\biggl[F_1(x_{1},x_{2})\biggr]; \qquad
F(z, z^{\star})=F_1(x_{1},x_{2}) + i F_2(x_{1},x_{2})
\label{fg}
\ee
(and, of course, vice versa). In order to find the particular solution of inhomogeneous
equation (\ref{FF}) one has to introduce new variables $y_1, y_2$ such that $y_2\equiv x_2$
and $y_1$ coincides with one of the solutions (\ref{fg}), e.g. $y_1\equiv F_1(x_1, x_2).$
In new variables Eq.(\ref{FF}) becomes:
\ba
\frac{\partial F_1(x_1, x_2)}{\partial x_1} \frac{\partial F_2(y_1, y_2)}{\partial y_2}=C,
\nonumber
\ea
and its solution for $F_2$ in terms of $F_1$ is:
\ba
F_2(x_1, x_2) = C\int dy_2 \Biggl( \frac{\partial F_1(x_1,x_2)}{\partial x_1}\Biggr)^{-1},
\nonumber
\ea
where the partial derivative is taken with $x_2=const$, and after integration over $y_2$
the initial variables $x_1,x_2$ are to be reinserted. So
the general solution of (\ref{FF}) is the sum:
\be
F_2(x_1,x_2) = \Phi\biggl[F_1(x_{1},x_{2})\biggr] +
C\int dy_2 \Biggl( \frac{\partial F_1(x_1,x_2)}{\partial x_1}\Biggr)^{-1}.
\label{gen}
\ee

Thus two matrix two-dimensional Hamiltonians (components of new
Superhamiltonian) $H^{(0)}$ (see expression (\ref{fac1})) and $(\widetilde H^{(1)}+C),$
which can be written as:
\ba
\widetilde H^{(1)} + C = H^{(0)} + 2\sigma_1\partial_1\partial_2F_1(x_1,x_2)
- 2\sigma_3\partial_1\partial_2F_2(x_1,x_2),
\nonumber
\ea
are intertwined by the
differential matrix operator (component of new supercharges):
\ba
Q^+ &\equiv& q^+\tilde q^+= \partial_1^2 -\partial_2^2
+2i\sigma_2\partial_1\partial_2 +\nonumber\\ &+&
\frac{1}{4}\biggl[ (\partial_2F_2)^2 + (\partial_2F_1)^2 - (\partial_1F_2)^2 -
(\partial_1F_1)^2 \biggr] +\nonumber\\ &+& \sigma_1 \biggl[(\partial_1F_1)\partial_2 -
(\partial_2F_1)\partial_1 - \frac{1}{2} (\partial_1^2+\partial_2^2)F_2 \biggr]
-\nonumber\\ &-& \frac{i}{2}\sigma_2 \biggl[ (\partial_1F_1)(\partial_2F_1) +
(\partial_1F_2)(\partial_2F_2) \biggr]
+ \nonumber\\ &+& \sigma_3 \biggl[ (\partial_2F_2)\partial_1 - (\partial_1F_2)\partial_2 +
\frac{1}{2} (\partial_1^2 +\partial_2^2)F_1\partial_1   \biggr]
\nonumber
\ea
The SUSY algebra has now a polynomial form:
\ba
\{\hat Q^+,\hat Q^-\} = \hat H (\hat H -C);\quad (\hat Q^+)^2 =(\hat
Q^-)^2 = 0; \quad [\hat H ,\hat Q^{\pm}] =0.
\nonumber
\ea
This algebra differs essentially from the deformed SUSY algebra
for a class of two-dimensional systems investigated in a series of papers
\cite{david} in so far as it {\bf does not} incorporate any non-trivial symmetry operators
(no central extension):
thus we have no positive indications of integrability of the systems $H^{(0)},\, \widetilde H^{(1)}.$

{\bf Higher order iterations.} In order to perform the next iteration of supersymmetric (Darboux)
transformations one has to glue $\widetilde H^{(1)}$ with the first
component $\widetilde{\widetilde{H}}^{(0)}$ of a new Superhamiltonian. This procedure
follows the previous one: superpotentials are expressed in
terms of derivatives of the same arbitrary complex function
$\widetilde F(z, z^{\star}):$
\ba
\widetilde W &=&\widetilde W_1 + i \widetilde W_2 =
+\bar\partial F(z, z^{\star}) =
-\partial \widetilde F; \label{tildeF}\\
\widetilde{\widetilde{W}}&=&\widetilde{\widetilde{W}}_1 + i \widetilde{\widetilde{W}}_2 =
+\bar\partial\widetilde F(z, z^{\star})     \label{ttildeF}.
\ea
The last equality in (\ref{tildeF}) implies that
in terms of an arbitrary function $ f(z, z^{\star}): $
\ba
F(z, z^{\star})=-\partial f(z, z^{\star});\qquad
\widetilde F(z, z^{\star})=+\bar\partial f(z, z^{\star}).
\nonumber
\ea
It is difficult to find the general solution of the nonlinear equations (\ref{1}) and
the analogous one for $\widetilde W,\,\widetilde{\widetilde W},$ but it is possible to
find particular solutions
with certain simplifying ansatzes. For example, explicit solutions can be
obtained for $C=\widetilde C =0$ and
$f(z,z^{\star})\equiv |\phi (z)|^2.$

\section*{\bf 3.\quad Iterations by gluing and global transformations}
\vspace*{0.2cm}
\hspace*{3ex} We now modify the procedure
(namely, Eq.(\ref{gluing})) of
the previous Section by gluing the Hamiltonians $H^{(1)}$ and $\widetilde H^{(0)}$
{\bf up to a global unitary transfomation} $U.$
Such a procedure allows to preserve the main details of the scheme,
but it might extend the class of intertwined Hamiltonians.

Without loss of generality, we may restrict
ourselves\footnote{From now on we will refer to such global unitary
transformation as "rotation" even disregarding $i.$} to $U= i\sigma_3 :$
\be
H^{(1)}=U\widetilde H^{(0)}U^{-1} + C, \qquad U=i\sigma_3.
\label{ggluing}
\ee
In this case (\ref{1}), (\ref{2}) must be replaced by:
\ba
|W(z, z^{\star})|^{2} &=& |\widetilde W(z, z^{\star})|^{2} + C; \label{11}\\
\partial_1(\widetilde W_2-W_2) &=& \partial_2(W_1+\widetilde W_1);\label{222}\\
\partial_1(\widetilde W_1+W_1) &=& \partial_2(W_2-\widetilde W_2).
\label{22}
\ea
The Eqs.(\ref{222}), (\ref{22}) can be solved explicitly:
$W$ and $\widetilde W$ are interrelated by an
arbitrary {\bf holomorphic function} $G(z):$
\be
W(z, z^{\star})+\widetilde{W}^{\star}(z, z^{\star})=G(z);\quad
G(z)\equiv G_{1}+iG_{2};\,\, \partial_{1}G_{1}=\partial_{2}G_{2};
\quad \partial_{1}G_{2}=-\partial_{2}G_{1}.
\label{G}
\ee
The remaining nonlinear equation (\ref{11}) for $ G $ can be
written as:
\be
G_{1}^{2} + G_{2}^{2} - 2 W_{2}G_{2} - 2 W_{1}G_{1} + C =0 .
\label{remaining}
\ee
It can be solved explicitly
as a linear {\bf algebraic equation} for superpotentials $ W_{1}, W_{2}$
in terms of an arbitrary holomorphic function $ G(z).$
For an arbitrary $G(z)$ one can choose also an arbitrary
function $W_2(z,z^{\star})$ and find the functions $W_1(z,z^{\star}),\,
\widetilde W_{1,2}(z,z^{\star})$ from Eqs. (\ref{remaining}), (\ref{G}).
Thus one obtains a class of Hamiltonians (\ref{fac1}) for which
the second order matrix supersymmetric transformations can be performed.
The resulting Hamiltonian
\be
\widetilde H^{(1)} + C = H^{(0)} +
2\sigma_1\cdot\biggl( \partial_1W_2-\partial_2W_1 +\partial_2G_1 \biggr)
+ 2\sigma_3\cdot\partial_1G_1
\label{HtildeH}
\ee
is intertwined with the initial Hamiltonian $H^{(0)}$ by the operator
\ba
Q^{+}=q^{+}\sigma_{3}\tilde q^{+},
\nonumber
\ea
and is therefore isospectral (up to zero modes of $ Q^{\pm} $) to $H^{(0)}.$

\hspace*{2ex} {\bf Example with hidden symmetry.} Among many possible examples we consider the simple peculiar case
when the resulting $\widetilde H^{(1)}$
{\bf coincides} with the initial $H^{(0)}$ up to a constant:
\be
H^{(0)} = \widetilde H^{(1)} + C.
\label{coincide}
\ee
This can be achieved by imposing (see (\ref{HtildeH}))
$G(z) \equiv iaz $ and
\ba
W_2(z,z^{\star})= \frac{1}{2}(a\rho +
\frac{C}{\rho})\cos \phi + \Theta(\rho)\sin \phi;
\quad W_1(z,z^{\star}) = \frac{1}{2}(a\rho +
\frac{C}{\rho})\sin \phi + \Theta(\rho)\cos \phi,
\nonumber
\ea
where $\rho , \phi$ are polar coordinates, $a$ is a constant and $\Theta(\rho)$ is an
arbitrary function. The Hamiltonian (\ref{coincide}) then becomes:
\be
H^{(0)} = -\Delta^{(2)} + \frac{1}{4} (a\rho + \frac{C}{\rho})^2 + (\Theta(\rho))^2 -
a\sigma_1 - \frac{1}{\rho}(\rho\Theta(\rho))^{\prime}\sigma_3.
\label{H1H0}
\ee
We stress that though the potential in (\ref{H1H0}) depends on $\rho$ only, its matrix structure
in general prevents a standard separation of variables, typical for scalar problems.
From the intertwining
(now commutation) relations $[H^{(0)}, Q^{\pm}] = 0$ one can find two mutually commuting
Hermitean symmetry operators:
\ba
R^- &\equiv &\frac{1}{2i}(Q^+ - Q^-) = ia\partial_{\phi};\label{R-}\\
R^+ &\equiv & \frac{1}{2}(Q^++Q^-) = -\sigma_3 H^{(0)} +ia\sigma_2\rho\partial_{\rho} +
\sigma_3\frac{1}{2}a\rho (a\rho +\frac{C}{\rho})
-\sigma_1 a\rho \Theta(\rho),
\label{R+}
\ea
with column-eigenfunctions, common to $H^{(0)}, R^{\pm}$:
\ba
e^{im\phi} \hat{\psi}_m(\rho).
\nonumber
\ea
The first one, $R^-,$ reflects the obvious symmetry of $H^{(0)}$ under rotations.,
but the second symmetry operator (\ref{R+}) realizes certain "hidden" symmetry and has
the property:
\ba
(R^+)^2 + (R^-)^2= H^{(0)}(H^{(0)} - C).
\nonumber
\ea
One can notice that the spectrum of the Hamiltonian involves double $m \leftrightarrow -m$ degeneracy, and
in addition a possible double degeneracy (for each value of $m$)
associated with a column-function $\hat{\psi}_m(\rho).$ A pedagogical example
illustrating the considerations above is obtained from (\ref{H1H0}) with
$\Theta (\rho)\equiv 0$ and $C=0:$ it is exactly solvable after
suitable rotation and leads to a pair of
decoupled radial oscillators with frequiency $a$ and a relative ground state
energy shift $2a.$

\hspace*{2ex} {\bf Higher order iterations.} In general, the procedure of gluing with a global rotation
enlarges essentially the class of Hamiltonians for which our method is applicable.
We now demonstrate this considering two options to construct {\bf third order
transformations:}

\hspace*{2ex} 1. {\bf The first option} reads:
\ba
H^{(0)} \longrightarrow H^{(1)}\equiv \widetilde H^{(0)} + C
\longrightarrow \widetilde H^{(1)} + C \equiv \sigma_3 \widetilde{\widetilde H}^{(0)}\sigma_3+C+
\widetilde C \longrightarrow
\widetilde{\widetilde H}^{(1)}+C+\widetilde C,
\nonumber
\ea
where $\longrightarrow$ denotes the appropriate first order matrix SUSY transformation
induced by $q^+, \tilde q^+, \tilde{\tilde q}^+,$ and $\equiv$ represents the gluing.
The first gluing is of the kind described in Section 2 (Eq.(\ref{gluing})), and
the superpotentials $W(z,z^{\star})$ and $\widetilde W(z,z^{\star})$
satisfy Eqs.(\ref{F}), (\ref{FF}). The second gluing is described in
Eq.(\ref{ggluing}), and
$\widetilde W(z,z^{\star}),\,\widetilde{\widetilde W}(z,z^{\star})$
satisfy Eqs.(\ref{G}), (\ref{remaining}).

Eq.(\ref{remaining}) can be rewritten as:
\be
\biggl( G(z)\partial F^{\star} +G^{\star}(z^{\star})\bar\partial F\biggr) -
|G|^2 - \widetilde C = 0,
\label{GF}
\ee
where $F(z,z^{\star})$ still satisfies (\ref{FF}). We now restrict ourselves to the case
$C=0,$ for which  Eq.(\ref{FF}) has trivial particular solutions with
purely real or purely imaginary $F$.
Then, e.g. for real $F$, Eq.(\ref{GF}) can be
solved by replacing $z,\,z^{\star}$ by:
\be
t\equiv \int\frac{dz}{2G(z)} + \int\frac{dz^{\star}}{2G^{\star}(z^{\star})};\quad
\tau\equiv i \Biggl( \int\frac{dz}{2G(z)} -
\int\frac{dz^{\star}}{2G^{\star}(z^{\star})}\Biggr),
\label{variables}
\ee
since Eq.(\ref{GF}) in new variables  reads:
\ba
\partial_t F_1 = |G|^2 + \widetilde C.
\nonumber
\ea
A solution  for $F_1$:
\ba
F_1 = T(\tau) + \int|G|^2dt + \widetilde Ct
\nonumber
\ea
depends on two arbitrary functions - one holomorphic $(G(z))$ and one real
$(T(\tau)).$ The $F=F_1$ so obtained generates $W$ and $\widetilde W$
when inserted into (\ref{F}) and (\ref{G}) (in terms of the initial variables $x_1,x_2$).
Thus one obtains a class of Hamiltonians
$H^{(0)},\, (\widetilde{\widetilde H}^{(1)}+C+\widetilde C), $ which are intertwined
(and are isospectral) by the third order
operator $Q^+ = q^+\tilde q^+\sigma_3\tilde{\tilde q}^+.$

\hspace*{2ex} 2. {\bf The second option} includes two gluings, both with $\sigma_3$-rotations:
\ba
H^{(0)} \longrightarrow H^{(1)}\equiv \sigma_3\widetilde H^{(0)}\sigma_3 + C
\longrightarrow \widetilde H^{(1)} + C \equiv \sigma_3 \widetilde{\widetilde
H}^{(0)}\sigma_3+C+\widetilde C \longrightarrow
\widetilde{\widetilde H}^{(1)}+C+\widetilde C.
\nonumber
\ea
The superpotentials $W(z,z^{\star})$ and $\widetilde W(z,z^{\star})$ has to be found from
equations of the form (\ref{remaining}), and then
$\widetilde W(z,z^{\star})$ and $\widetilde{\widetilde W}(z,z^{\star})$
from equations of the form (\ref{G}). One can obtain the general solution
of these equations
purely algebraically in terms of two arbitrary holomorphic functions $G(z),\,
\widetilde G(z):$
\be
W(z,z^{\star}) = \frac{G^2\widetilde G - |\widetilde G|^2 G - \widetilde C G - C \widetilde G^{\star}}
{\widetilde G G - \widetilde G^{\star} G^{\star}},
\label{WWW}
\ee
$\widetilde W(z,z^{\star}),\, \widetilde{\widetilde W}(z,z^{\star})$ being derived,
again algebraically, from
two equations of form (\ref{G}). Thus one obtains a new class of Hamiltonians
$H^{(0)},\, (\widetilde{\widetilde H}^{(1)}+C+\widetilde C), $ which are intertwined by the third order
$Q^+ = q^+\sigma_3\tilde q^+\sigma_3\tilde{\tilde q}^+,$ and are therefore
isospectral up to zero modes of $Q^{\pm}.$

\section*{\bf 4.\quad Second order supersymmetrization of scalar two-dimensional Hamiltonians.}
\vspace*{0.2cm}
\hspace*{3ex}
In this concluding Section we will discuss particular cases
of Sections 2 and 3, when the initial Hamiltonian $ H^{(0)} $
is constrained to be diagonal. The condition of diagonality of (\ref{fac1}) reads:
\be
\partial_{2}W_{1}-\partial_{1}W_{2} = 0  \quad \Leftrightarrow\quad
W_{2}=\partial_{2}\chi (x_{1},x_{2});\qquad W_{1}=
\partial_{1}\chi (x_{1},x_{2}),
\label{condition}
\ee
where $ \chi (x_{1},x_{2}) $ is an arbitrary {\bf real} function.
In terms of this function one has $ H^{(0)}=diag \bigl( H^{(0)}_{11},\, H^{(0)}_{22}\bigr) $
with
\be
H^{(0)}_{11} =
-\Delta^{(2)}+(\partial_{i}\chi)^2 -
(\partial_{1}^{2}+\partial_{2}^{2})\chi; \quad
H^{(0)}_{22} = H^{(0)}_{11} +2(\partial_{1}^{2} +\partial_{2}^{2})\chi ,
\label{h22}
\ee
which is intertwined
(and therefore, isospectral, up to zero modes of supercharges (\ref{q}))
with non-diagonal
Hamiltonian $ H^{(1)}. $ The latter is obtained by inserting
(\ref{condition}) into (\ref{fac2}).
From now on we study the iteration of SUSY transformations for the initial
Hamiltonian (\ref{h22}), i.e. the compatibility of conditions
(\ref{condition}) with the iteration algorithm.

\hspace*{2ex} {\bf Simple gluing.}
For the case $C=0$ in (\ref{gluing}) the combination of Eq.(\ref{F}) with Eq.(\ref{condition}) can be
solved in terms of an arbitrary real function
$ \xi (x_{1},x_{2}) : $
\ba
F_{1}(x_{1},x_{2})&=&(\partial_{2}^{2}-\partial_{1}^{2})\xi (x_{1},x_{2});
\label{xi1}\\
F_{2}(x_{1},x_{2})&=& - 2 \partial_{1}\partial_{2}\xi (x_{1},x_{2});
\label{xi2}\\
\chi (x_{1},x_{2})&=&(\partial_{1}^{2}+\partial_{2}^{2})\xi (x_{1},x_{2}).
\label{xi3}
\ea
The last condition to be taken into account is the interrelation
(\ref{fg}) between $ F_{1} $ and $ F_{2} :$
\be
(\partial_{2}^{2}-\partial_{1}^{2})\xi (x_{1},x_{2}) =
f \bigl[\partial_{1}\partial_{2}\xi (x_{1},x_{2})\bigr].
\label{ff}
\ee
Of course, one can not find the general solution of this nonlinear functional-differential
equation, but one can find a variety of particular
solutions. The simplest choice $ f=const $ in (\ref{ff})
is trivial, since the
corresponding Hamiltonian $ H^{(1)} $ can be directly diagonalized by
rotation. Some less trivial solutions of (\ref{xi1}),
(\ref{xi2}), (\ref{ff}) can be found by requiring
that the l.h.s. of Eqs.(\ref{xi1}), (\ref{xi2}) are e.g. $ F_{i}=\lambda_{i}\xi^k;
\quad F_{i}=\exp{\lambda_{i}\xi}, \,$ etc.

For the case $C \neq 0$ one can try similar techniques as used for the solution
of Eq.(\ref{FF}) to solve now the equation (\ref{FF}) and (\ref{condition})
together with (\ref{F}). The resulting equation reads:
\ba
(\partial_1^2 - \partial_2^2) F_2 = 2\partial_1\partial_2 F_1.
\nonumber
\ea
Particular solutions of this equation can be found by suitable ansatzes like,
for example:
\ba
F_1(x_1,x_2) \equiv x_1 g(x_2) + f_2(x_2);\quad \Phi (F_1) \equiv a F_1^2 + b F_1 +
c + \frac{d}{F_1},
\nonumber
\ea
where $g(x_2), f(x_2)$ can be determined by solvability conditions, the functional $\Phi (F_1)$
has been introduced in (\ref{gen}), and $a,b,c,d$ are constants. One can also enlarge
the class of solutions by suitable nonsingular linear transformations of $x_1,x_2.$

\hspace*{2ex} {\bf Gluing with rotation.}
In this case the nonlinear equation (\ref{remaining}) takes the form:
\ba
(\partial_{1}\chi ) G_{1} +
(\partial_{2}\chi ) G_{2} =
1/2 \bigl[ G_{1}^{2} + G_{2}^{2} + C\bigr],
\nonumber
\ea
or, in terms of variables $z, z^{\star}:$
\be
\bigl( G(z)\partial + G^{\star}(z^{\star})\bar\partial\bigr) \chi (z,z^{\star}) = |G(z)|^2 +
\frac{1}{4}C.
\label{rrremaining}
\ee
It has the form similar to Eq.(\ref{GF}) of the previous Section except for the fact
that $\chi (z,z^{\star})$ is a real function. Therefore, the general solution of
(\ref{rrremaining}) is expressed  in terms of $t,\,\tau$ of (\ref{variables})
via an arbitrary (real) function $\Lambda (\tau):$
\be
\chi (z,z^{\star}) = \Lambda (\tau) + \int|G(z)|^2dt + \frac{1}{4}Ct.
\label{lambda}
\ee
The initial diagonal Hamiltonian $H^{(0)}$ from (\ref{fac1}) and (\ref{condition}) reads:
\ba
H^{(0)} &=& -\Delta^{(2)} + 4(\partial\chi)(\bar\partial\chi) - 4(\partial\bar\partial\chi)
\cdot\sigma_3 =\nonumber\\
&=& -\Delta^{(2)} +|G|^2 + \frac{1}{2}C + \frac{C^2}{16|G|^2} +
\frac{\Lambda^{\prime 2}(\tau)}{|G|^2} +
\frac{2\Lambda^{\prime}(\tau)}{|G|^2}\int\partial_{\tau}|G|^2dt +
\frac{1}{|G|^2}\Bigl(\int\partial_{\tau}|G|^2dt\Bigr)^2 -\nonumber\\
&-&
\sigma_3\cdot\Bigl( G^{\prime} + G^{\star\prime} + \frac{\Lambda^{\prime\prime}}{|G|^2} \Bigr),
\nonumber
\ea
while the final $\widetilde H^{(1)}$ can be expressed as:
\ba
\widetilde H^{(1)} + C = H^{(0)} - 2\sigma_1 \partial_1G_2 + 2\sigma_3 \partial_1G_1.
\nonumber
\ea
As a particular example one can consider the case of $G = z^{1/2}$ which leads to a
confining singular Hamiltonian with a singularity $1/\rho$ and
growing asymptotically as $\rho$ with a nontrivial azimuthal dependence.

Finally, we note that conversely the intertwining between diagonal
$ H^{(0)} $ and non-diagonal $\widetilde H^{(1)}$ by second order matrix supercharges
can also be used to diagonalize a class of matrix
Schr\"odinger operators (or system of differential equations of
second order with non-diagonal matrix coupling), in analogy with
similar procedures already used in \cite{pauli1}, \cite{pauli2}, \cite{channels}.

\section*{\bf Acknowledgements}
\vspace*{0.2cm}
\hspace*{3ex}
M.V.I. and D.N.N. are grateful to the University of Bologna and INFN for
support and kind hospitality.
This work was partially supported by the Russian Foundation for Fundamental
Research (Grant No.02-01-00499).


\vspace{.2cm}

\begin{thebibliography}{}
\bibitem{witten} E. Witten 1981 {\it Nuclear Physics} {\bf B185} 513
\bibitem{review} G. Junker 1996 {\it Supersymmetric Methods in Quantum
and Statistical
Physics} (Springer, Berlin)\\
F. Cooper, A. Khare, U. Sukhatme 1995 {\it Phys. Rep.} {\bf 25} 268\\
B.K. Bagchi 2001 {\it Supersymmetry in Quantum and Classical Mechanics}
(Chapman and Hall/CRC, New York)
\bibitem{ais} A.A. Andrianov, M.V Ioffe, V.P. Spiridonov 1993 {\it Phys.Lett.}
{\bf A174} 273
\bibitem{acdi} A.A. Andrianov, F. Cannata, J.-P. Dedonder, M.V. Ioffe 1995
{\it Int.J.Mod.Phys.} {\bf A10} 2683
\bibitem{HSUSY}
B.F. Samsonov 1996 {\it Mod. Phys. Lett.} {\bf A11} 1563\\
D.J. Fernandez C., M.L. Glasser, L.M. Nieto 1998 {\it Phys.Lett.} {\bf A240} 15\\
V.G. Bagrov, B.F. Samsonov, L.A. Shekoyan 1998 quant-ph/9804032\\
S. Klishevich, M. Plyushchay 1999 {\it Mod. Phys. Lett.} {\bf A14} 2739\\
A.A. Andrianov, F. Cannata, M.V. Ioffe, D.N.Nishnianidze 2000
{\it Phys.Lett.} {\bf A266} 341\\
D.J. Fernandez C., J. Negro, L.M. Nieto 2000 {\it Phys.Lett.} {\bf A275} 338\\
S. Klishevich, M. Plyushchay 2001 {\it Nucl.Phys.} {\bf B606[PM]} 583\\
H. Aoyama, M. Sato, T. Tanaka 2001 {\it Phys.Lett.} {\bf B503} 423\\
H. Aoyama, M. Sato, T. Tanaka 2001 {\it Nucl.Phys.} {\bf B619} 105\\
R. Sasaki, K. Takasaki 2001 {\it J. Phys. A: Math.Gen.} {\bf 34} 9533\\
A.A. Andrianov, A.V. Sokolov 2003 {\it Nucl.Phys.} {\bf B660} 25
\bibitem{darboux} G. Darboux 1882 {\it Compt. Rend.} {\bf 94} 1456
\bibitem{calogero} F. Calogero, A. Degasperis {\it Spectral Transform
and Solitons} {v.1} (North-Holland Publishing Company, Amsterdam)
\bibitem{crum} M.M. Crum 1955 {\it Quart. J. Math. Oxford} {\bf 6} 121
(reproduced by H.Rosu in physics/9908019)\\
M.G. Krein 1957 {\it Dokl. Akad. Nauk SSSR} {\bf 113} 970 (in Russian)
\bibitem{matrix}
A.A. Andrianov, F. Cannata, M.V. Ioffe, D.N.Nishnianidze 1997
{\it J. Phys. A: Math.Gen.} {\bf 30} 5037
\bibitem{samsonov} B.F. Samsonov, A.A. Pecheritsin 2003 quant-ph/0307145
\bibitem{abi}
A.A. Andrianov, N.V. Borisov, M.V. Ioffe 1984 {\it JETP Lett.}
{\bf 39} 93\\
A.A. Andrianov, N.V. Borisov, M.V. Ioffe 1984 {\it Phys. Lett.}
{\bf A105} 19\\
A.A. Andrianov, N.V. Borisov, M.V. Ioffe 1985 {\it Theor. Math.Phys.}
{\bf 61} 1078
\bibitem{abei} A.A. Andrianov, N.V. Borisov, M.V. Ioffe, M.I. Eides 1985
{\it Phys. Lett.} {\bf 109A} 143\\
A.A. Andrianov, N.V. Borisov, M.V. Ioffe, M.I. Eides 1984
{\it Theor. Math. Phys.} {\bf 61} 965
\bibitem{pauli1}
A.A. Andrianov, M.V. Ioffe 1988 {\it Phys. Lett.} {\bf B205} 507
\bibitem{pauli2}
M.V. Ioffe, A.I. Neelov 2003 {\it J. Phys. A: Math.Gen.} {\bf 36} 2493
\bibitem{david}  A. Andrianov, M. Ioffe, D. Nishnianidze 1995
{\it Phys.Lett.}, {\bf A201} 103\\
A.A. Andrianov, M.V. Ioffe, D.N. Nishnianidze 1995 {\it Theor. Math. Phys.}
{\bf 104} 1129\\
A.A. Andrianov, M.V. Ioffe, D.N. Nishnianidze 1996 solv-int/9605007;
Published in: 1995 {\it Zapiski Nauch.
Seminarov POMI RAN} ed.L.Faddeev et.al. {\bf 224} 68 (In Russian);
Translation in:
{\it Problems in QFT and Statistical Physics} ed.L.D.Faddeev
et.al. {\bf 13}\\
A.A. Andrianov, M.V. Ioffe, D.N. Nishnianidze 1999
{\it J.Phys.:Math.Gen.} {\bf A32} 4641
\bibitem{two} F. Cannata, M.V. Ioffe, D.N. Nishnianidze 2002
{\it J.Phys.:Math.Gen.} {\bf A35} 1389
\bibitem{kamran} A. Gonzalez-Lopez, N. Kamran 1998
{\it J.Geom.Phys.} {\bf 26} 202
\bibitem{mateos} A. Alonso Izguierdo, M.A. Gonzalez Leon,
M. de la Torre Mayado, J. Mateos Guilarte 2004
hep-th/0401054
\bibitem{para}  A. Andrianov, M. Ioffe, V.P. Spiridonov, L.Vinet 1991
{\it Phys.Lett.}, {\bf B272} 297
\bibitem{channels} F. Cannata, M.V. Ioffe 1993 {\it J.Phys.:Math.Gen.} {\bf A26} L89

\end{thebibliography}
\end{document}